\documentclass[10pt,journal]{IEEEtran}

\usepackage{graphicx}          
\usepackage{amsmath}           
\usepackage{amssymb}
\usepackage{algorithm}
\usepackage{algpseudocode}
\usepackage{epstopdf}
\usepackage{cite}
\usepackage{url}
\usepackage{subcaption}
\usepackage[utf8]{inputenc}  

\usepackage{tikz}
\usetikzlibrary{positioning, calc, shapes.geometric, shapes, shapes.multipart, arrows.meta, arrows, decorations.markings, external, trees}
\tikzstyle{Arrow} = [
	thick, 
	decoration={
		markings,
		mark=at position 1 with {
			\arrow[thick]{latex}
			}
		}, 
	shorten >= 3pt, preaction = {decorate}
]

\usepackage{array}
\usepackage{booktabs}
\usepackage{multirow}
\usepackage{siunitx}

\usepackage{amsthm}

\theoremstyle{definition}

\theoremstyle{remark}

\theoremstyle{plain}

\theoremstyle{definition}

\DeclareMathOperator{\E}{\mathbb{E}}

\DeclareMathOperator*{\argmin}{arg\,min}

\newcommand{\mbf}[1]{\mathbf{#1}}
\newcommand{\mbs}[1]{\boldsymbol{#1}}
\newcommand{\what}[1]{\widehat{#1}}
\newcommand{\wtilde}[1]{\widetilde{#1}}

\newcommand{\0}{\mbf{0}}

\newcommand{\T}{\top}

\newcommand{\x}{\mbs{x}}

\newcommand{\z}{\mbs{z}}

\newcommand{\w}{\mbs{w}}

\newcommand{\basisfunc}{\mbs{\phi}}
\newcommand{\wopt}{\mbs{w}_o}
\newcommand{\wconst}{\mbs{w}_c}
\newcommand{\wopthat}{\what{\mbs{w}}_o}
\newcommand{\wconsthat}{\what{\mbs{w}}_c}
\newcommand{\ztail}{\mathcal{Z}}
\newcommand{\vtheta}{\mbs{\theta}}

\newcommand{\cmat}{\mbs{\Sigma}}

\newcommand{\zpredx}{\mbs{\Gamma}}
\newcommand{\xpredy}{\mbs{\alpha}}
\newcommand{\zpredy}{\mbs{\beta}}

\newcommand{\MSE}{\text{MSE}}
\newcommand{\MSEtail}{\text{MSE}_\alpha}
\newcommand{\MSEtyp}{\text{MSE}_{1-\alpha}}
\newcommand{\projnull}{\mbs{\Pi}}
\newcommand{\wset}{\mathcal{W}}

\begin{document}

\title{Robust Prediction when Features are Missing}
\author{Xiuming~Liu, 
        Dave~Zachariah, 
        Petre~Stoica,~\IEEEmembership{Fellow,~IEEE}%
\thanks{Manuscript submitted 23 December, 2019. This research has been partly supported by the Swedish Research Council via the projects 2017-04543, 2017-04610, and 2018-05040. The authors are with the Department of Information Technology, Uppsala University, Sweden.}
}

\maketitle

\begin{abstract}
Predictors are learned using past training data which may contain features that are unavailable at the time of prediction. We develop an approach that is robust against outlying missing features, based on the optimality properties of an oracle predictor which observes them. The robustness properties of the approach are demonstrated on both real and synthetic data.
\end{abstract}

\section{Introduction}
A common task in statistical machine learning and signal processing is to predict an outcome $y$ based on features $\x$ and $\z$, using past training data drawn from an unknown distribution
$$(\x_i, \z_i, y_i) \sim p( \x, \z, y), \quad i=1,\dots, n.$$
In certain problems, however, not all features in the training data are available at the time of prediction \cite{saar2007handling,anava2015online,mercaldo2018missing}. For instance, in medical diagnosis certain features are more expensive or time-consuming to obtain than others, and therefore unavailable in an early stage of assessment \cite{little2012prevention}. Other features are  observable only after the outcome has occurred. We let $\z$ denote the features \emph{missing} at the time of prediction and consider the task of predicting $y$ given only the observable features $\x$.

A direct approach predicts only on the basis of the association between $\x$ and $y$, and thus discard all past training data containing $\z$. By contrast, missing data in statistics is commonly tackled by means of imputation \cite{rubin1996multiple,schafer2000inference,chapelle2006semi,saar2007handling,little2019statistical}. An indirect approach is then to predict $y$ using both $\x$ and a regression imputed $\what{\z}(\x)$. These two approaches, however, turn out to be equivalent, as we explain in Section~\ref{sec:problem}. We show that linearly parameterized predictors that minimize the mean squared error (MSE) result in high prediction errors when the missing features are outliers. That is, when $\z$ occurs far from its mean in the test sample.

Robust statistics has typically focused on outlier problems in contaminated training data \cite{huber1964robust,lange1989robust, christmas2010robust}. Robust learning of model parameters is then often achieved by considering regression models with t-distributed noise. The focus of this paper, however, is robust prediction in the case of outlying missing features. Specifically, the aim is to reduce the large errors in the outlier case, while incurring a minor increase of errors for the inlier cases. This is particularly relevant in applications with real losses, such as predicting health-related outcomes which are associated with future covariates that are unavailable in the test sample.


 We achieve robustness using an adaptively weighted combination of optimistic and conservative predictors, which are derived in Section~\ref{sec:predictors}. The approach of switching between modes during extreme events can be found in econometrics \cite{hamilton1994autoregressive} and signal processing \cite{poritz1982linear,ephraim2005revisiting}, but has not been considered for prediction with missing features. We demonstrate the robustness properties of the proposed approach using both synthetic and real data sets.

\emph{Notation:} We let $\| \x \|_{\mbf{W}} = \sqrt{\x^\T \mbf{W} \x}$, where $\mbf{W} \succeq \0$, and define the sample mean of  $\x$ as $\E_n[\x] = n^{-1}\sum_{i=1}^{n}\x_i$. The pseudoinverse of a matrix $\mbs{A}$ is denoted by $\mbs{A}^\dagger$.

\begin{figure}
    \centering
    \includegraphics[width = .35\textwidth]{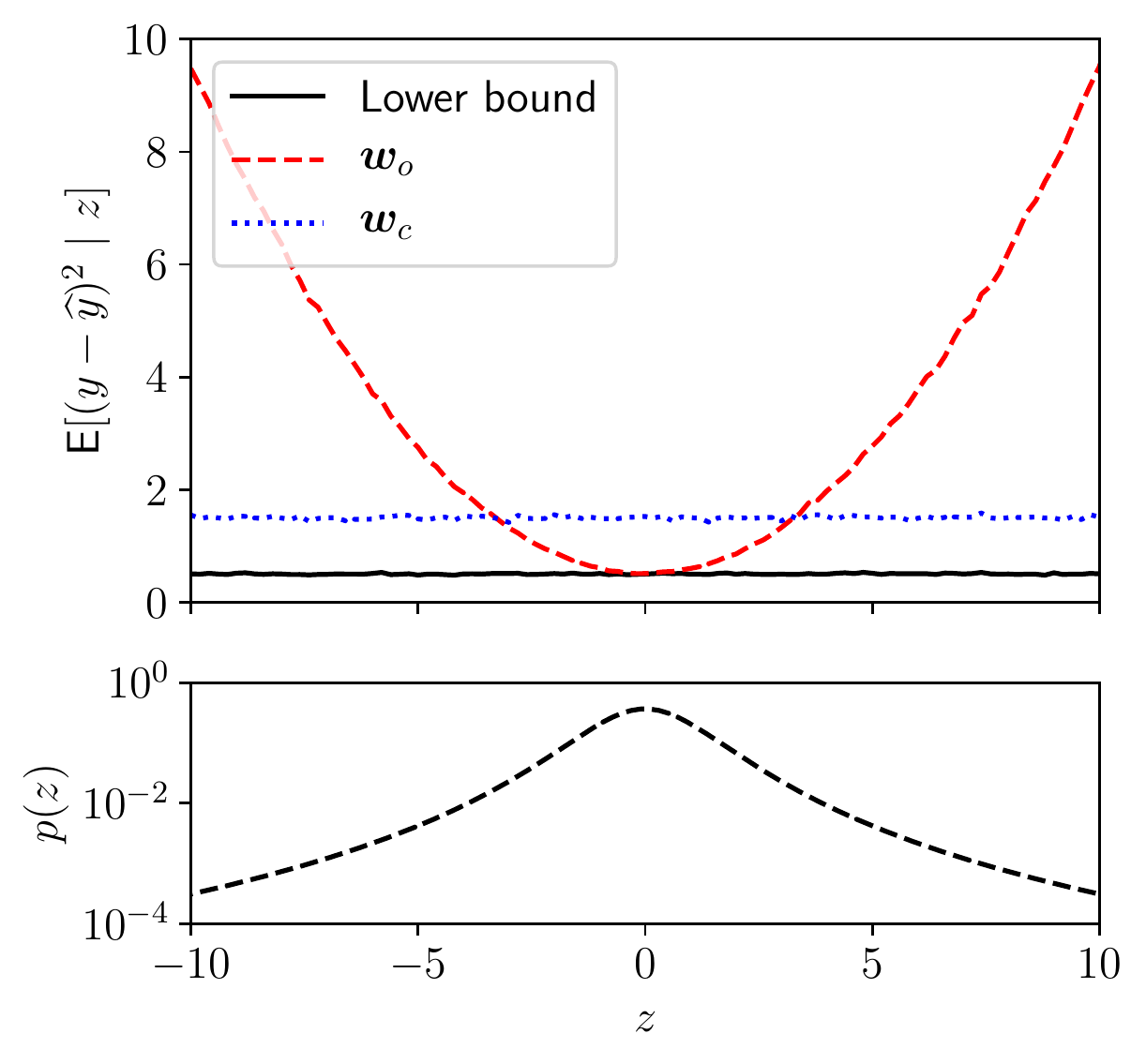}
    \caption{Illustration of MSE conditioned on a missing scalar feature $z$ along with $p(z)$. The optimistic predictor $\wopt$ may lead to high outlier $\MSEtail$ for the tails of $p(z)$. By contrast, the conservative predictor $\wconst$ (see Section \ref{sec:predictors} for definition) can mitigate the outlier events. The MSE is lower bounded by that of an oracle predictor which observes $z$. The example used to generate this figure is specified in Section \ref{sec:synthetic}.}
    \label{fig:example1}
\end{figure}

\section{Problem formulation}
\label{sec:problem}

We consider scenarios in which: 1) $\x$ and $\z$ are correlated; 2) the dimension of $\x$ is greater than that of $\z$. 
We study the class of linearly parameterized predictors $\what{y}(\x; \w) = \w^\T \x$, where $\w \in \mathbb{R}^d$. Without loss of generality we consider $(\x, \z, y)$ to be centered. Note that the results in this paper can be readily extended to arbitrary functions of the features by replacing $\x$ in $\w^\T \x$ with a function $\basisfunc(\x)$. 

The mean squared-error of a predictor is
\begin{equation}
\MSE(\w) \triangleq \E\left[ \big|y - \what{y}(\x; \w)\big|^2 \right],
\label{eq:mse}
\end{equation}
where the expectation is with respect to $(\x, \z, y)$. In the rest of this section we discuss briefly how a missing feature $\z \in \mathbb{R}^q$ affects the prediction performance. The tails of the distribution of $\z$ are contained in the region
\begin{equation}
    \ztail_\alpha = \Big\{ \z : \z^\T (\E\left[\z \z^\T\right])^{-1} \z \geq q/\alpha \Big \}
\label{eq:tailregion}
\end{equation}
as $\alpha$ approaches 0. Indeed $\Pr\{ \z \in \ztail_\alpha \} \leq \alpha$ (see the appendix), and thus a small $\alpha$ corresponds to the probability of an outlier event. Making use of $\ztail_\alpha$, we can decompose the MSE into an outlier $\MSEtail = \E\big[ |y - \what{y}|^2 \; | \; \z \in \ztail_\alpha\big]$ and an inlier $\MSEtyp = \E\big[ |y - \what{y}|^2 \; | \; \z \not \in \ztail_\alpha\big]$:
$$\MSE = \Pr\{ \z \in \ztail_\alpha \} \MSEtail + \Pr\{ \z \not \in \ztail_\alpha \} \MSEtyp.$$
As Figure~\ref{fig:example1} illustrates, the prediction performance can degrade significantly for outlier events.

Using $n$ training samples, our goal is to formulate a robust predictor that will reduce the outlier $\MSEtail$ without incurring a significant increase of the inlier $\MSEtyp$.

\section{Predictors: Optimistic, Conservative and Robust}
\label{sec:predictors}

If the feature $\z$ were known, the optimal linearly parameterized predictor would be given by
\begin{equation}
\what{y}_\star(\x, \z) = \xpredy^\T\x + \zpredy^\T \z,
\label{eq:oracle}
\end{equation}
where
\begin{equation}
\begin{bmatrix} \xpredy \\ \zpredy \end{bmatrix} = \begin{bmatrix}\E[\x\x^\T] &  \E[\x\z^\T]\\ 
\E[\z\x^\T] &  \E[\z\z^\T]\end{bmatrix}^{-1}\begin{bmatrix}\E[\x y] \\  \E[\z y]\end{bmatrix}
\label{eq:oracleparameters}
\end{equation}
Its prediction errors are uncorrelated with both sets of features, that is,
\begin{equation}
\boxed{\E\left[ \x(y - \what{y})\right] = \0 \quad \text{and} \quad  \E\left[ \z(y - \what{y})\right] = \0,}
\label{eq:orthogonality}
\end{equation}
which renders the predictor robust against outlier events for both $\x$ and $\z$, respectively. In the case of missing features $\z$, we begin by  considering predictors $\what{y}(\x; \w) = \w^\T \x$ which satisfy either one of the orthogonality properties in \eqref{eq:orthogonality}.

\subsection{Optimistic predictor}
The predictors that satisfy the first equality in \eqref{eq:orthogonality} are given by the parameter vectors in
\begin{equation}
\wset_o = \left\{ \w : \E\left[ \x ( y - \w^\T \x ) \right] = \0 \right\} 
\label{eq:weightsetopt}
\end{equation}
This set, however, consists of a single element $\wopt$ which is also the minimizer of \eqref{eq:mse}\cite{kailath2000linear}. That is,
\begin{equation}
\wopt \equiv \argmin_{\w} \: \MSE(\w) = (\E[\x \x^\T])^{-1}\E[\x y]
\label{eq:wopt}
\end{equation}
We denote the resulting predictor as `optimistic' with respect to the missing $\z$, because it does not attempt to satisfy the second equality in \eqref{eq:orthogonality}, see Fig.~\ref{fig:example1} for an illustration of its performance.

\emph{Remark:} Given that $\x$ and $\z$ are correlated, we may consider using the MSE-optimal linear predictor 
\begin{equation}
    \what{\z} = \E[\z \x^\T](\E[\x \x^\T])^{-1} \x 
\label{eq:zpred}
\end{equation}
to impute the missing feature. An indirect predictor approach would then be to use $\what{\z}$ in \eqref{eq:oracle} in lieu of $\z$, but this is equivalent to \eqref{eq:wopt}. That is, $\what{y}_\star(\x, \what{\z}) \equiv \wopt^\T \x = \what{y}(\x; \wopt)$, which can be shown by using the block matrix inversion lemma in \eqref{eq:oracleparameters}.

\subsection{Conservative predictor}
The predictors that satisfy the second equality in \eqref{eq:orthogonality} are given by all parameter vectors in
\begin{equation}
\wset_c = \left\{ \w : \E\left[ \z ( y - \w^\T \x ) \right] = \0 \right\} 
\label{eq:weightsetconst}
\end{equation}
This set is a $(d-q)$-dimensional subspace of $\mathbb{R}^d$ and therefore we can consider the parameter vector that minimizes \eqref{eq:mse}, viz.
\begin{equation}
\wconst = \argmin_{\w \in \wset_c  } \: \MSE(\w),
\label{eq:wconst}
\end{equation}
We denote the resulting predictor as `conservative' with respect to the missing $\z$, because it satisfies only the second equality in \eqref{eq:orthogonality}, see Fig.~\ref{fig:example1} for an illustration of its performance.

\emph{Remark:}
Comparing the error of $\what{y}(\x; \w)$ with that of $\what{y}_\star(\x, \z)$ in \eqref{eq:oracle}, the excess MSE can be expressed as
\begin{equation}
\begin{split}
&\MSE(\w) - \MSE{}_\star \\
&= \| \zpredx(\xpredy - \w) + \zpredy \|^2_{\E[\z \z^\T]}+ \| \xpredy - \w \|^2_{\E[\wtilde{\x} \wtilde{\x}^\T]} \; \geq \; 0,
\end{split}
\label{eq:excessmse}
\end{equation}
where $\zpredx = (\E[\z \z^\T])^{-1}\E[\z \x^\T]$ and $\widetilde{\x} = \x - \zpredx^\T \z$ is a residual (see the appendix). Note that the first term in \eqref{eq:excessmse} is weighted by the dispersion of $\z$. The constraint $\w \in \wset_c$ forces this term to zero. This leaves $d-q$ degrees of freedom that can be used to minimize the second term. By contrast, $\wopt$ minimizes the sum of both terms.

\subsection{Robust predictor}
Satisfying only one of the equalities in \eqref{eq:orthogonality} comes at a cost: The optimistic $\wopt$ yields robustness against outlying $\x$ but not $\z$ and, conversely, the conservative $\wconst$ yields robustness against outlying $\z$ but not $\x$. Since both equalities can be satisfied only by the infeasible predictor \eqref{eq:oracle}, we propose a predictor that interpolates between the optimistic and conservative modes using the side information that $\x$ provides about outliers in the missing features $\z$. That is, we propose to learn the adaptive parameter vector
\begin{equation}
\boxed{\w(\x) =   \Pr\{ \z \not \in \ztail_\alpha | \x \} \wopt + \Pr\{ \z \in \ztail_\alpha | \x \} \wconst,}
\label{eq:strategy}
\end{equation}
such that the predictor $\what{y}(\x) = \w^\T(\x) \x$ becomes robust against outliers in both $\x$ and $\z$.

\section{Learning the robust predictor}
Learning the robust predictor requires finding  finite-sample approximations of $\wopt$, $\wconst$ and $\Pr\{ \z \in \ztail_\alpha | \x \}$ in \eqref{eq:strategy} using $n$ training samples $\{ (\x_i, \z_i, y_i )\}$.

We begin by defining $\MSE_n(\w) = \E_n[|y-\w^{\T} \x|^2]$, which yields the empirical counterpart of \eqref{eq:wopt}:
\begin{equation}
\wopthat = \argmin_{\w} \: \MSE_n(\w) = \big(\E_n\left[\x \x^\T\right]\big)^{\dagger}\E_n[\x y]
\label{eq:wopt_hat}
\end{equation}
Note that the pseudoinverse is used to include cases in which the sample covariance matrices are degenerate. Similarly, for \eqref{eq:wconst} we note that the empirical counterpart of the constraint in \eqref{eq:weightsetconst} is
\begin{equation}
    \E_n[\z ( y - \w^\T \x ) ] = \mbs{0} \Leftrightarrow  \E_n\left[ \z \x^\T \right] \w = \E_n[\z y]
\label{eq:constraint_emp}
\end{equation}
All vectors that satisfy \eqref{eq:constraint_emp} can therefore be parameterized as
\begin{equation*}
    \w(\vtheta) = (\E_n\left[ \z \x^\T \right])^\dagger \E_n\left[ \z y \right] + \projnull \vtheta,
\end{equation*}
where $\projnull$ is the orthogonal projection matrix onto the null space of $\E_n\left[ \z \x^\T \right]$ and $\vtheta \in \mathbb{R}^{d-q}$ is arbitrary. This yields the empirical counterpart of \eqref{eq:wconst}
\begin{equation}
\wconsthat = \w(\what{\vtheta}),
\label{eq:wconst_hat}
\end{equation}
where $\what{\vtheta}$ is the minimizer of the quadratic function $\MSE_n(\w(\vtheta))$.

\begin{figure}
	\centering
	\includegraphics[width=.33\textwidth]{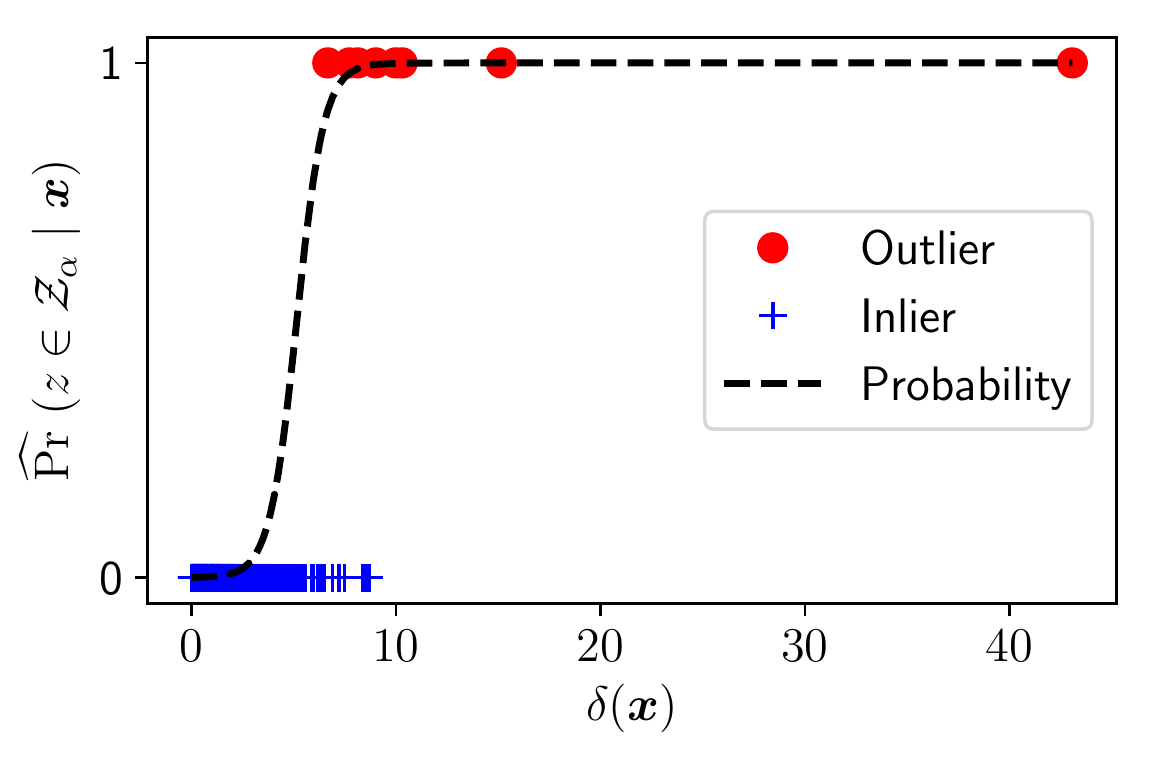}
	\caption{Outlier and inlier $\z$ in the training data (for $\alpha = 10\% $) versus statistic $\delta(\x)$, along with fitted logistic model \eqref{eq:logisticmodel} of outlier probability conditioned on $\x$ . In this example, the learned parameters were $\delta_0 = 4.21$ and $\kappa = -1.78$, respectively. The example used to generate this figure is specified in Section \ref{sec:synthetic}.}
	\label{fig:logistic}
\end{figure}

Next, we consider learning a model of the probability of an outlier event, $\Pr\{ \z \in \ztail_\alpha | \x \}$, conditioned on $\x$. Noting the definition \eqref{eq:tailregion}, we predict an outlier event using the scalar
$$\delta(\x) = \sqrt{\what{\z}^\T(\x) \E_n[\z \z^\T]^\dagger  \what{\z}(\x)  } \; \geq \; 0,$$
where $\what{\z}(\x) =   \E_n[\z \x^\T](\E_n[\x \x^\T])^{\dagger} \x$  is the empirical version  of \eqref{eq:zpred}. The conditional outlier probability is modeled using a standard logistic function,
\begin{equation}
\what{\Pr}\{ \z \in \ztail_\alpha | \x \} = \frac{1}{1 + \exp \kappa(  \delta(\x)  - \delta_0)}
\label{eq:logisticmodel}
\end{equation}
The model parameters $\kappa$ and $\delta_0$ are learned from the training data $\{ (\x_i, \z_i) \}$ by minimizing the standard cross-entropy criterion
\begin{equation}
\begin{split}
\min_{\kappa, \delta_0}&\; -\E_n\Big[ I(\z \in \ztail_\alpha)  \ln \what{\Pr}\{ \z \in \ztail_\alpha | \x \} \\
&\qquad + I(\z \not \in \ztail_\alpha) \ln (1- \what{\Pr}\{ \z \in \ztail_\alpha | \x \}) \Big].
\end{split}
\label{eq:crossentropy}
\end{equation}
This approach takes into account the inherent uncertainty of predicting an outlying $\z$ from $\x$. An example of a fitted model as in \eqref{eq:crossentropy} is presented in Figure \ref{fig:logistic}. In practice, the targeted level $\alpha$ must not be too low in order to include a sufficient number of outlying training samples.

In sum, we learn a robust predictor with an adaptive parameter vector
\begin{equation}
\what{\w}(\x) = \what{\Pr}\{ \z \not \in \ztail_\alpha | \x \} \wopthat + \what{\Pr}\{ \z \in \ztail_\alpha | \x \} \wconsthat
\label{eq:strategy_finite}
\end{equation}
using $n$ samples, as described in Algorithm~\ref{alg:summary}.
\begin{algorithm}
\caption{Learning the robust predictor}
\label{algorithm}
\begin{algorithmic}[1]
\State \textbf{Input:} Training data $\{( \x_i, y_i, \z_i )\}$ and $\alpha \in (0,1]$
\State Compute $\wopthat$ via \eqref{eq:wopt_hat} \State Compute $\wconsthat$ via \eqref{eq:wconst_hat}
\State For each $(\x_i,\z_i)$, form $(\delta(\x_i), I(\z_i \in \ztail_\alpha))$ 
\State Learn $\what{\Pr}\{ \z \in \ztail_\alpha | \x \}$ via \eqref{eq:crossentropy}
\State \textbf{Output:} $\what{\w}(\x)$ in \eqref{eq:strategy_finite} 
\end{algorithmic}
\label{alg:summary}
\end{algorithm}

\emph{Remark:} In the case of high-dimensional features $\x$ and $\z$ one may use regularized  methods, such as ridge regression, \textsc{Lasso}, or the tuning-free \textsc{Spice} method \cite{zachariah2015online}, to learn $\wopthat$, $\wconsthat$ and $\what{\z}(\x)$.

\section{Experimental results}
We evaluate the robustness of the proposed predictor using both synthetic and real data. Additional results are in the supplementary material. The data and code for reproducing the experiment results are available at: https://github.com/xiumingliu/robust-regression.git. 

\begin{figure*}
    \centering
    \begin{subfigure}[b]{0.3\textwidth}
        \includegraphics[width=\textwidth]{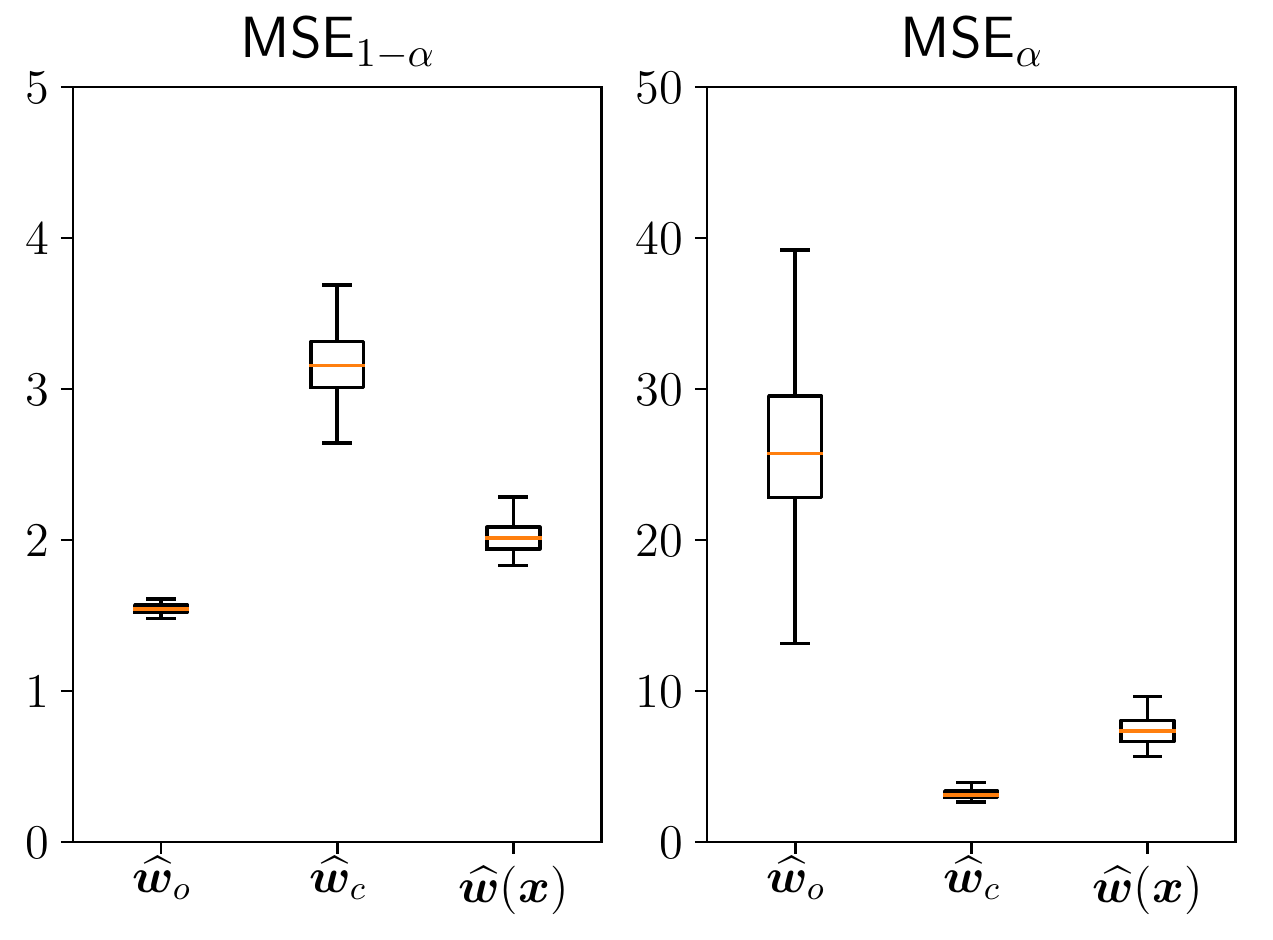}
        \caption{$\rho = 0.7$}
    \end{subfigure}
    \hfill
    \begin{subfigure}[b]{0.3\textwidth}
        \includegraphics[width=\textwidth]{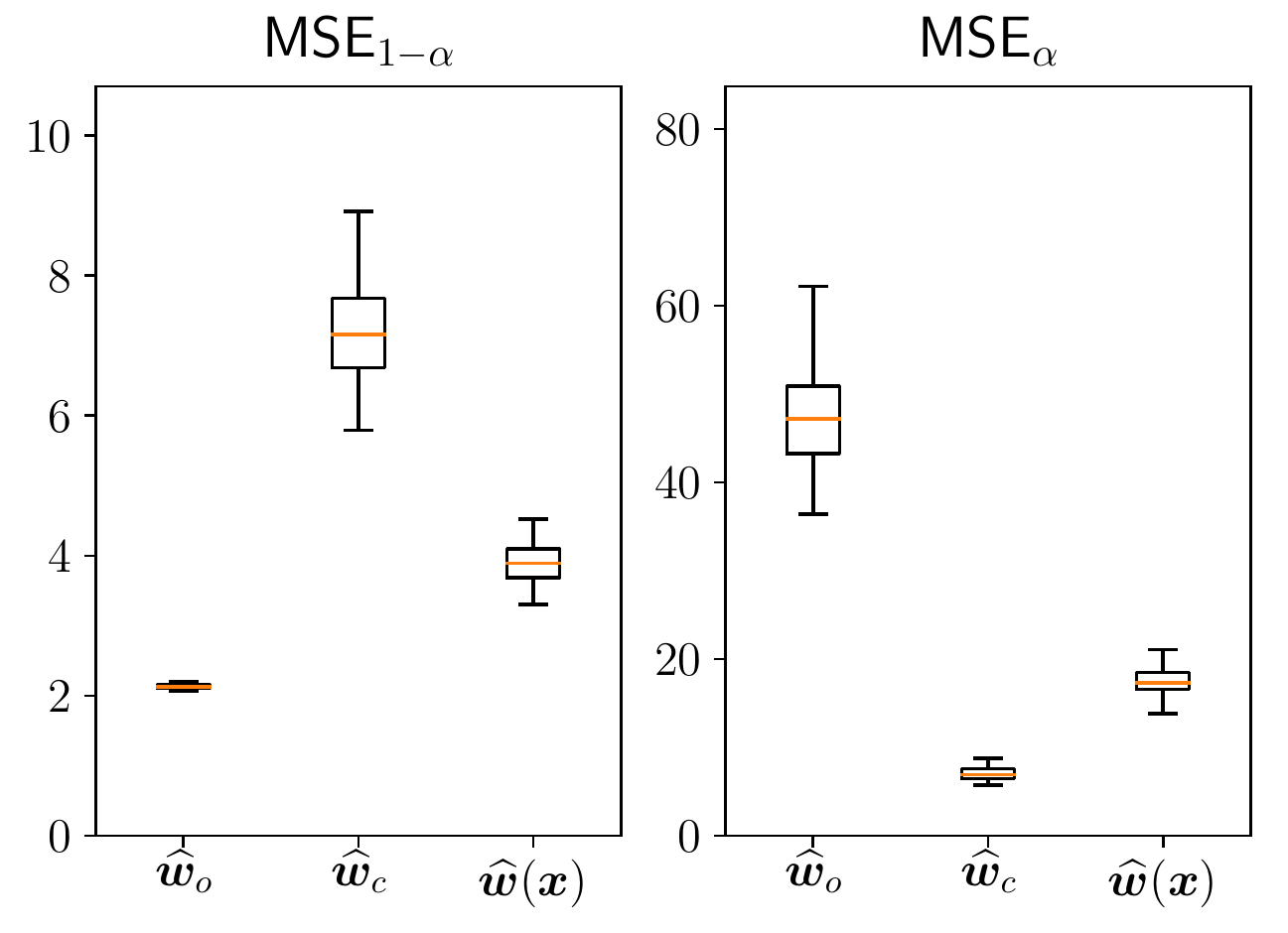}
        \caption{$\rho = 0.5$}
    \end{subfigure}
    \hfill
    \begin{subfigure}[b]{0.3\textwidth}
        \includegraphics[width=\textwidth]{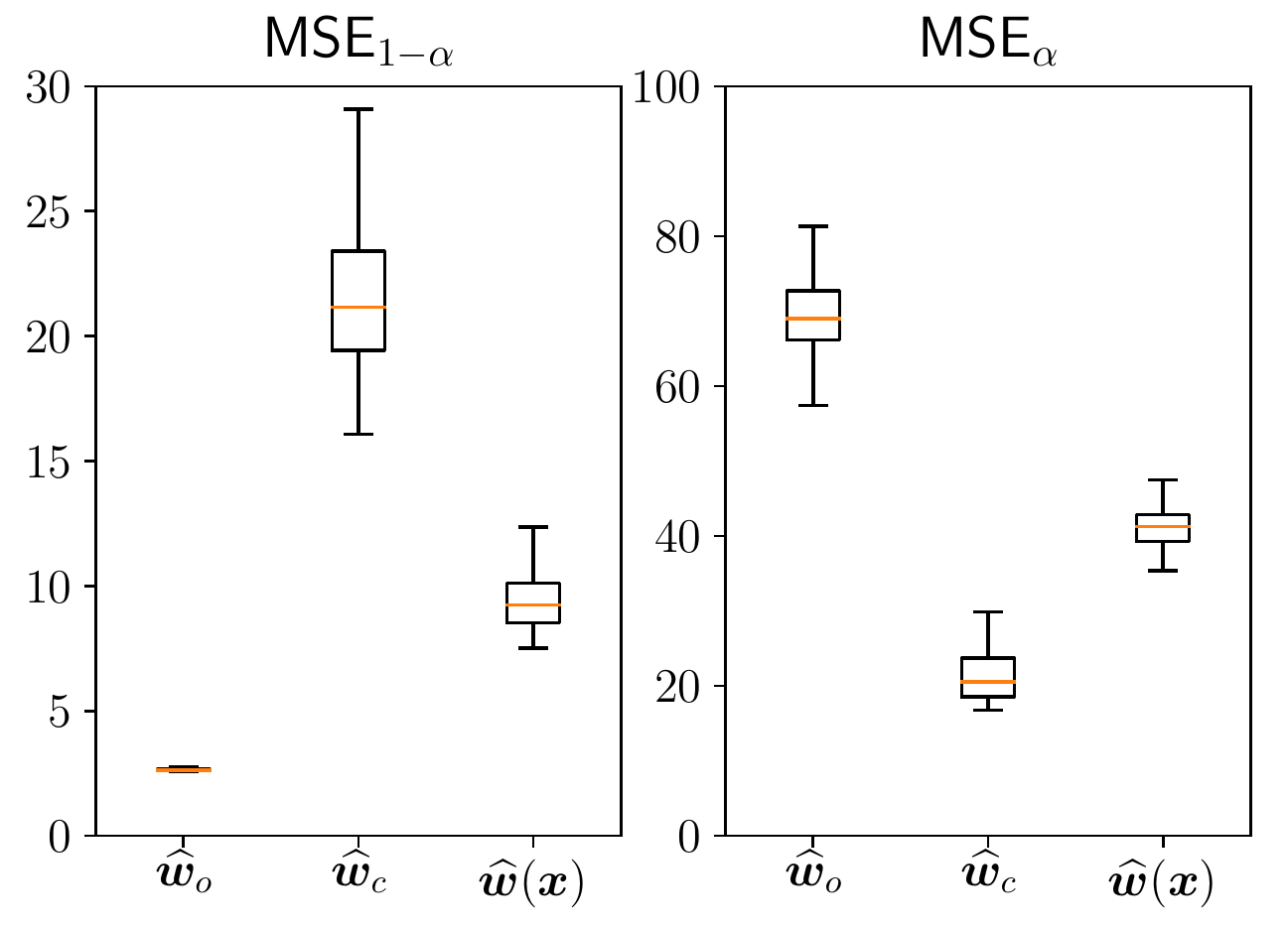}
        \caption{$\rho = 0.3$}
    \end{subfigure}
    \caption{The box plot of $\MSEtyp$ and $\MSEtail$ for $\alpha = 10\%$ and different $\rho$. The boxes show the distribution of outlier $\MSEtyp$ and inlier $\MSEtail$ from 50 simulations using $n = 100$ training and $10^6$ test samples. }
    \label{fig:different_rho}
\end{figure*}

\subsection{Synthetic data}
\label{sec:synthetic}
Consider the following data-generating process: $z \sim t(0, 1, \nu_z)$ and
\begin{equation}
\begin{split}
\x &=  \mbs{1}\rho z + \mbs{u} + \epsilon_{\x}  \: \in \mathbb{R}^3, \\
y &= z + \mbs{1}^\T \x + \epsilon_{y} \: \in \mathbb{R}, 
\end{split}
\label{eq:t_data_generate}
\end{equation}
where $\rho \in [-1,1]$ equals the correlation coefficient between $z$ and the elements of $\x$. The latent variable $\mbs{u} \sim t(\0, \cmat_{\mbs{u}}, \nu_{\mbs{u}})$ is t-distributed with $\nu_{\mbs{u}}$ degrees of freedom and $(\epsilon_{\x}, \epsilon_{y})$ are white Gaussian processes of corresponding dimensions.   

\begin{figure}
\centering
\includegraphics[width = .35\textwidth]{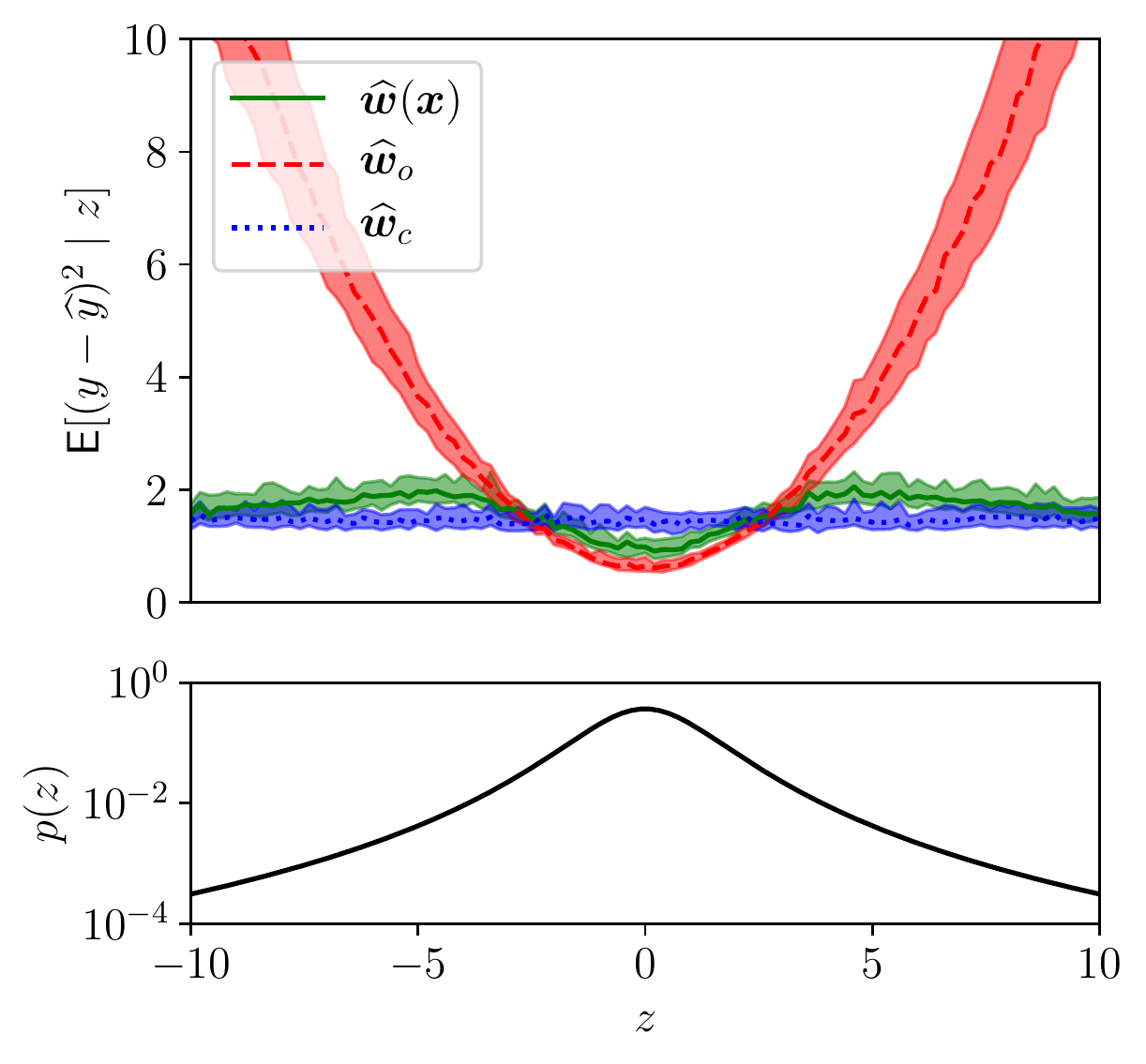}
\caption{Conditional MSE versus the missing feature $z$ over 50 MC runs, where the bands correspond to the $75\%$-percentiles. For $\alpha=10\%$, the outlier region $\ztail_\alpha$ contains the tails in which the magnitude of $z$ exceeds approximately $5.5$. 
In comparison with the optimistic $\wopt$, the robust predictor $\what{\w}(\x)$ significantly reduces the errors for outlier $z$, while incurring only a small increased error for inlier $z$.}
\label{fig:fat_tailed}
\end{figure}

We evaluate the predictors $\what{y}(\x; \w)$ of a new outcome $y$ given only $\x$, where the vector $\w$ is learned from $n$ samples of training data. Specifically, we evaluate the optimistic $\wopthat$, conservative $\wconsthat$ and proposed $\what{\w}(\x)$ predictors in a case of missing features with heavy tails ($\nu_{z} = 3$), where $n$ ranges from 100 to 1000 and $\alpha = 10\%$. A comparison of the conditional MSE functions of the learned predictors ($n = 1000$) is given in Fig.~\ref{fig:fat_tailed}, where it is seen that the robust predictor interpolates between the two modes.

When averaging over $\z$, the distributions of MSEs are illustrated in Fig.~\ref{fig:different_rho} using 50 Monte Carlo (MC) runs with different training data. We see that the robust predictor drastically reduces the outlier $\MSEtail$, while yielding only a small increase in the inlier $\MSEtyp$. The differences in MSEs, when averaged over all training datasets, are summarized in Tables \ref{table:fat_tailed}, which demonstrates the robustness of the proposed approach.


\begin{table}[hbt!]
    \centering
    \begin{tabular}{lllll}
    \toprule
        \multirow{2}{*}{$n$} & \multicolumn{2}{c}{$\wconsthat$} & \multicolumn{2}{c}{$\what{\w}(\x)$} \\ 
         & $\Delta \MSEtyp$ & $\Delta \MSEtail$ & $\Delta \MSEtyp$ & $\Delta \MSEtail$ \\ \midrule
        100 & +102\% & -74\% & +34\% & -63\% \\ 
        500 & +100\% & -86\% & +29\% & -70\% \\ 
        1000 & +103\% & -87\% & +30\% & -71\% \\ 
        \bottomrule
    \end{tabular}
    \caption{Changes in averaged MSE in comparison to the MSE of $\wopthat$, for $\alpha = 10\%$ and $\rho = 0.7$. Average over 50 MC runs using $10^6$ test samples. }
    \label{table:fat_tailed}
\end{table}

\subsection{Air quality data}
Next, we demonstrate the proposed method using real-world air quality data. Nitrogen-oxides (NO$_\text{x}$) emitted by the fossil fuel vehicles are major air pollutants in urban environments, with negative impacts on the health of inhabitants. 

The aim here is to predict the daily average of NO$_\text{x}$ concentration, denoted $y$, based on NO$_\text{x}$ and ozone (O$_3$) measurements from $L$ previous days. That is, $\x$ is of dimension $d=2L$ and contains the daily average NO$_\text{x}$ and O$_3$ levels from the $L$ past days. In the training data we have also access to $z$, the O$_3$ concentration, at the same time as the outcome $y$. This feature $z$ is correlated with $y$ and $\x$. For the prediction of a new outcome, however, $z$ is a  missing feature.

The dataset contains 10 years of daily average NO$_\text{x}$ and O$_3$ measurements from 2006-01-01 to 2015-12-31. Data is split into 7 years of training data (2006-2012), and 3 years of test data (2012-2015). We use $\alpha = 30\%$ to obtain sufficient outlier events in the training data for the model fitting \eqref{eq:crossentropy}. The prediction errors summarized in Table \ref{tabel:nox} show that we are able to reduce the outlier $\MSEtail$ by about 10\% while incurring a minimal increase of the inlier $\MSEtyp$.

\begin{table}[hbt!]
    \centering
    \begin{tabular}{lllll}
    \toprule
        \multirow{2}{*}{$L$} & \multicolumn{2}{c}{$\wconsthat$} & \multicolumn{2}{c}{$\what{\w}(\x)$} \\ 
         & $\Delta \MSEtyp$ & $\Delta \MSEtail$ & $\Delta \MSEtyp$ & $\Delta \MSEtail$ \\ \midrule
        7 & +4.1\% & -18.0\% & +0.7\% & -6.9\% \\ 
        28 & +3.5\% & -24.7\% & +0.4\% & -11.4\% \\ 
        56 & +3.1\% & -23.1\% & +0.4\% & -11.2\% \\ 
        \bottomrule
    \end{tabular}
    \caption{NO$_\text{x}$ prediction. Changes in the MSE in comparison to the MSE of $\wopthat$, for $\alpha = 30\%$.}
    \label{tabel:nox}
\end{table}

\section{Conclusion}
Based on the orthogonality properties of an optimal oracle predictor, we developed a predictor that is robust against outliers of the missing features. The proposed predictor is formulated as a convex combination of  optimistic and conservative predictors, and requires only specifying the intended outlier level against which it must be robust. The ability of the robust predictor to suppress outlier errors, while incurring only a minor increase in the inlier errors, was demonstrated using both simulated and real-world datasets.

\appendix
\section{}
\subsection{Probability bound for \eqref{eq:tailregion}}
The probability bound for an event $\z \in \ztail_\alpha$ follows readily from a Chebychev-type inequality:
\begin{equation*}
\begin{split}
\Pr\{ \z \in \ztail_\alpha \} 
&\leq  \int_{\ztail_\alpha} \left[ (\alpha/q) \z^\T (\E[\z \z^\T])^{-1} \z \right] p(\z) \: d\z \\
&\leq \alpha \int \left[  \z^\T (\E[\z \z^T])^{-1} \z /q \right] p(\z) \: d\z \\
&= \alpha
\end{split}
\end{equation*}

\subsection{MSE decomposition \eqref{eq:excessmse}}
The outcome $y$ can always be decomposed as 
\begin{equation}
y = \xpredy^\T\x + \zpredy^\T \z + v,
\label{eq:decomp}
\end{equation}
where $\MSE{}_\star = \E[v^2]$. The random variable $v$ is orthogonal to any linear function of $\x$ and $\z$; which includes the residual $\wtilde{\mbs{\x}} = \x - \zpredx^\T \z$. Using \eqref{eq:decomp} we express the prediction error as
\begin{equation*}
\begin{split}
y - \w^\top \x 
= [\zpredx(\xpredy - \w) + \zpredy]^\T\z + (\xpredy - \w)^\T\wtilde{\mbs{\x}} + v 
\end{split}
\end{equation*}
Squaring this expression and taking the expectation yields \eqref{eq:excessmse}. Similarly, inserting it into the constraint in \eqref{eq:weightsetconst} yields $\zpredx(\xpredy - \w) + \zpredy = \0$.

\subsection{Multivariate Polynomial Regression}
Consider the following data-generating process: $z \sim t(0, 1, \nu_z)$ and
\begin{equation}
\begin{split}
\x &=  \mbs{1}\rho z + \mbs{u} + \epsilon_{\x}  \: \in \mathbb{R}^3, \\
y &= \w^\T_z\psi(z) + \w^\T_x \phi(\x) + \epsilon_{y} \: \in \mathbb{R}, 
\end{split}
\label{eq:t_data_generate_poly}
\end{equation}
where $\rho \in [-1,1]$ equals the correlation coefficient between $z$ and the elements of $\x$. The latent variable $\mbs{u} \sim t(\0, \cmat_{\mbs{u}}, \nu_{\mbs{u}})$ is t-distributed with $\nu_{\mbs{u}}$ degrees of freedom and $(\epsilon_{\x}, \epsilon_{y})$ are white Gaussian processes of corresponding dimensions.   

We consider
\begin{align}
    \psi(z) =& \begin{bmatrix}z & z^2\end{bmatrix}^\T \\
    \phi(\x) =& \begin{bmatrix}x_1 & x_2 & x_3 & x_1^2 & x_2^2 & x_3^2\end{bmatrix}^\T, 
\end{align}
in the data generation process \eqref{eq:t_data_generate_poly} which corresponds to a multivariate polynomial regression model. The quadratic terms of the t-distributed variables follow the F-distribution. For the variances exist for the quadratic terms, we set $\nu_z = 5$ and $\nu_{\mbs{u}} = 5$. Let $\w_z = \begin{bmatrix} w_0 & w_1 \end{bmatrix}^\T$, where $w_1$ parameterizes the nonlinear effect of the missing variable $z$.

\begin{figure}
    \centering
    \begin{subfigure}[b]{0.23\textwidth}
        \includegraphics[width=\textwidth]{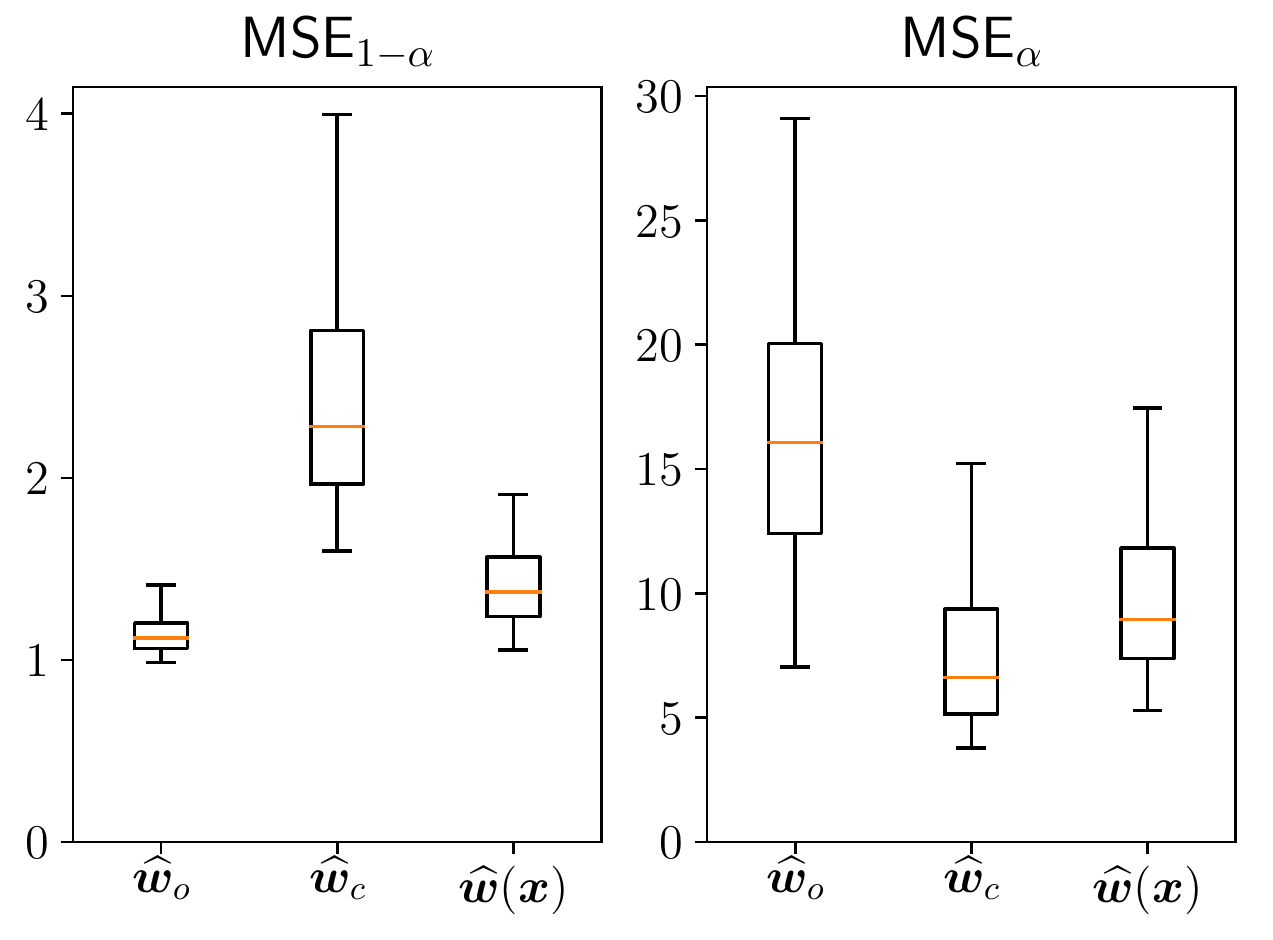}
        \caption{Nonlinear predictor, $w_1 = 0.1$}
    \end{subfigure}
    \hfill
    \begin{subfigure}[b]{0.23\textwidth}
        \includegraphics[width=\textwidth]{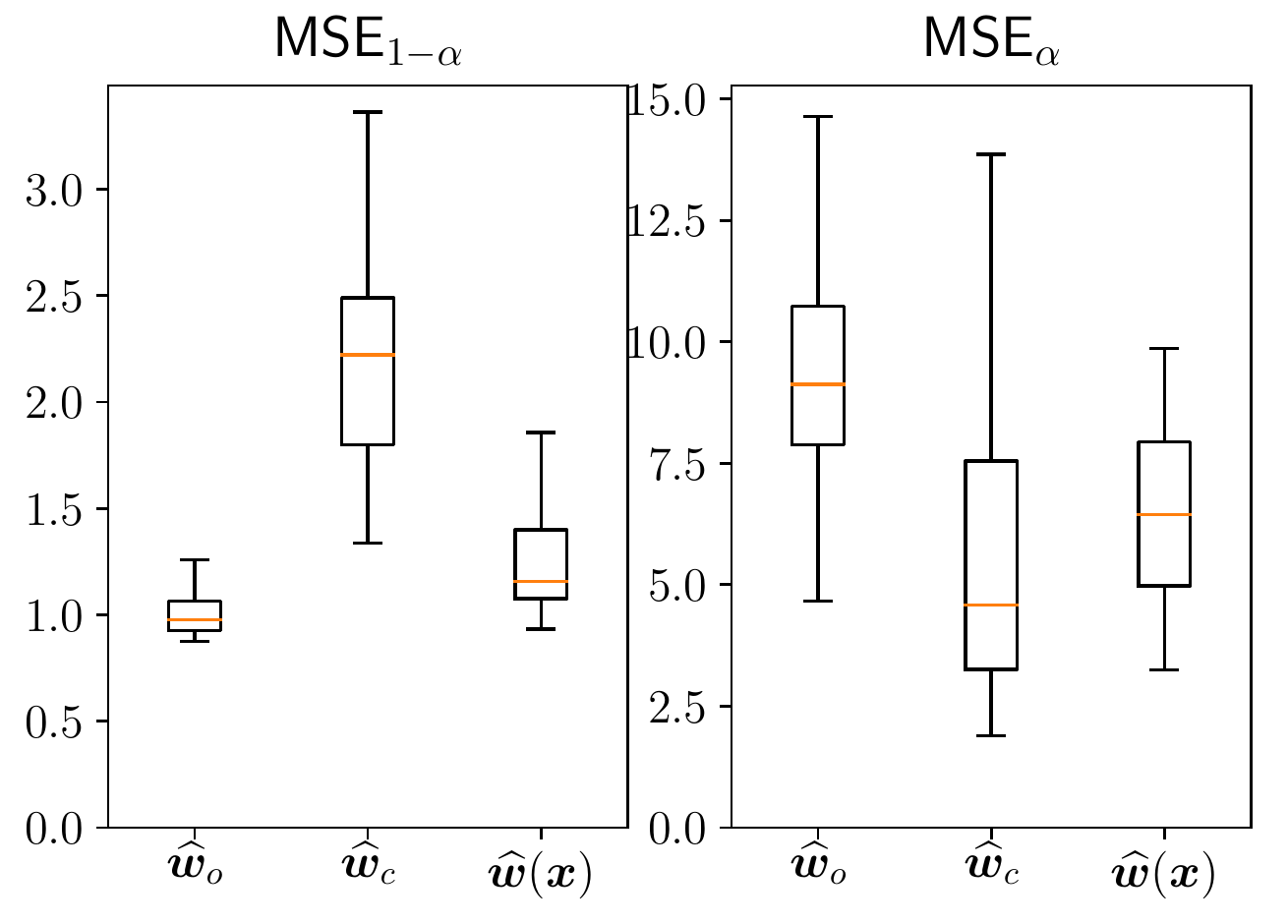}
        \caption{Nonlinear predictor, $w_1 = 0.01$}
    \end{subfigure}
    \\
    \begin{subfigure}[b]{0.23\textwidth}
        \includegraphics[width=\textwidth]{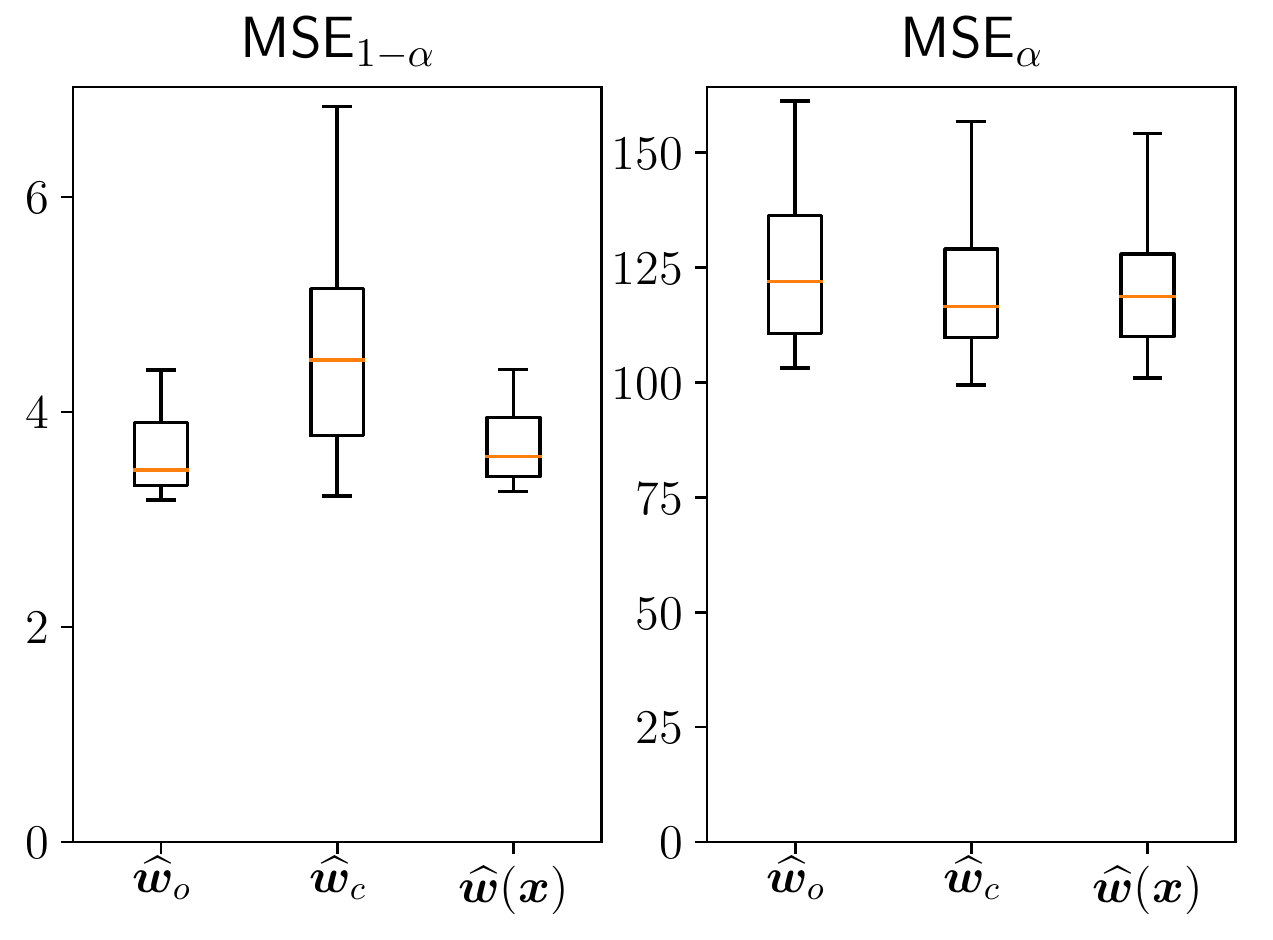}
        \caption{Linear predictor, $w_1 = 0.1$}
    \end{subfigure}
    \hfill
    \begin{subfigure}[b]{0.23\textwidth}
        \includegraphics[width=\textwidth]{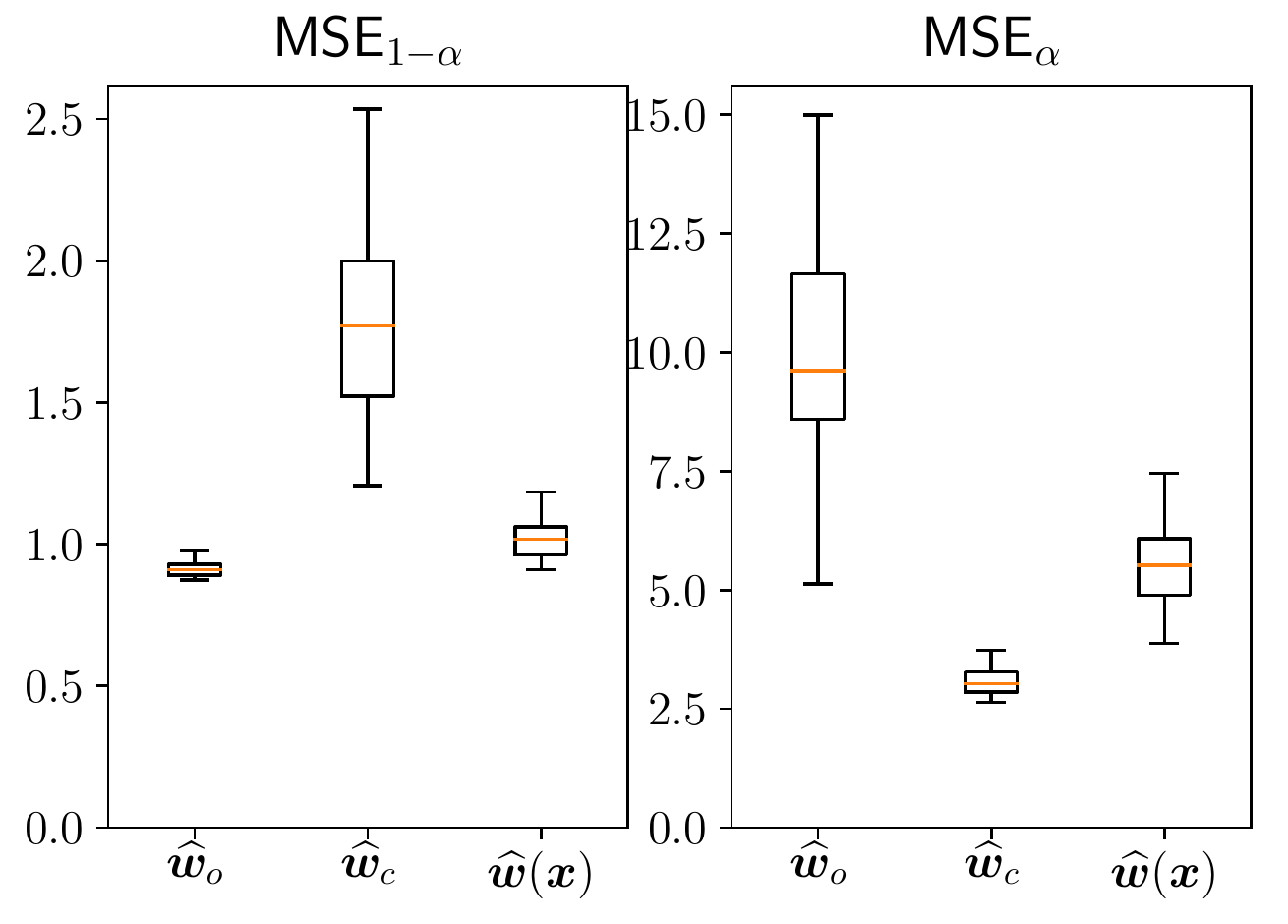}
        \caption{Linear predictor, $w_1 = 0.01$}
    \end{subfigure}
    \caption{Applying nonlinear and linear predictors to the data generated from process \eqref{eq:t_data_generate_poly}, with an unknown weight $w_1$ begin large (left column) and small (right column), respectively. }
    \label{fig:nonlinear_vs_linear}
\end{figure}

We apply the robust prediction method to the data using a nonlinear predictor of the form $\what{y}(\x) = \w^\T \phi(\x)$, treating $\psi(z)$ as missing features. For comparison, included the results when using a linear predictor $\what{y}(\x) = \w^\T \x$ and $z$ is missing. The results summarized in Figure \ref{fig:nonlinear_vs_linear} show that when the nonlinearity of the process is sufficiently large, the nonlinear predictor leads to dramatic reductions in outlier MSE. For negligible nonlinearities, the gains are nonexistent, as expected. In all cases, however, the proposed robust method leads to reduced outlier MSE, with a minor increase in inlier MSE.

\subsection{Diabetes data}
In this section we demonstrate the proposed method on the diabetes dataset \cite{efron2004least}. In the dataset, there are records from 442 patients. A record in the dataset consists of 10 features: age, sex, body mass index (BMI), average blood pressure (BP) and six blood serum measurements. The variable of interest is a  quantitative measure of disease progression one year after baseline. We divided the data into the training (size of 100) and the testing (size of 342) data, and assumed that one of the features (BMI or BP) is missing in the testing data. The changes of prediction errors (\%) are summarized in Table \ref{tabel:diabetes}. When BMI is missing in the testing samples, we reduce the outlier $\MSEtail$ by about 10\% while incurring a 5\% increase of the inlier $\MSEtyp$; when BP is missing, we reduce the outlier $\MSEtail$ by about 10\% while incurring only a 0.34\% increase of the inlier $\MSEtyp$. 

\begin{table}[hbt!]
    \centering
    \begin{tabular}{lSSSS}
    \toprule
        \multirow{2}{*}{Missing feature} & \multicolumn{2}{c}{$\wconsthat$} & \multicolumn{2}{c}{$\what{\w}(\x)$} \\ 
         & $\Delta \MSEtyp$ & $\Delta \MSEtail$ & $\Delta \MSEtyp$ & $\Delta \MSEtail$ \\ \midrule
        BMI & 17.94\% & -17.75\% & 5.73\% & -9.95\% \\ 
        BP & 2.45\% & -19.48\% & 0.34\% & -10.49\% \\
        \bottomrule
    \end{tabular}
    \caption{Diabetes response prediction. Changes in the MSE in comparison to the MSE of $\wopthat$, for $\alpha = 30\%$.}
    \label{tabel:diabetes}
\end{table}

\bibliographystyle{ieeetr}
\bibliography{reference}

\end{document}